\documentstyle[a4,11pt]{article}
\begin{document}

\title{{\bf A Gauge-invariant Analysis of Magnetic Fields in General
Relativistic Cosmology}}
\author{Christos G. Tsagas\thanks{%
e-mail address: c.tsagas@sussex.ac.uk}  and John D. Barrow\thanks{%
e-mail address: j.d.barrow@sussex.ac.uk} \\
%EndAName
{\small {\it Astronomy Centre, University of Sussex,}}\\
{\small {\it \ Brighton BN1 9QH, U.K.}}}
\maketitle

\begin{abstract}
We provide a fully general-relativistic treatment of cosmological
perturbations in a universe permeated by a large-scale primordial magnetic
field using the Ellis-Bruni gauge-invariant formalism. The exact non-linear
equations for general relativistic magnetohydrodynamic evolution are
derived. A number of applications are made: the behaviour of small
perturbations to Friedmann universes are studied; a comparison is made with
earlier Newtonian treatments of cosmological perturbations and some effects
of inflationary expansion are examined. 
\end{abstract}

\section{Introduction}

%~~~~~~~~~~~~~~~~~~~~~
The study of cosmological magnetic fields has many motivations. Any
primordial magnetic field could provide a seed field for dynamo
amplification in disc galaxies, and fields of order $10^{-10}$ gauss would
create significant magnetic fields in galaxy clusters through adiabatic
compression alone, \cite{PS}-\cite{ZH}. It would also introduce new
ingredients into the standard, but necessarily uncertain, picture of the
very early universe. Large-scale magnetic fields introduce anisotropies into
the expansion dynamics. They would not survive an epoch of inflation
although it is conceivable that large-scale fields and magnetic
inhomogeneities with a constant curvature spectrum could be generated at the
end of inflation. A number of proposals of this sort have been made, \cite
{TW}-\cite{Do}. They involve speculative changes to the nature of the
electromagnetic interactions in the universe at the time of inflation which,
as yet, have no other motivation other than to allow magnetic field
generation to take place. While such changes are not altogether
unreasonable, and may reveal important aspects of fundamental physics, it is
hard to test them independently. Ways of generating magnetic fields at
cosmological phase transitions have also been explored by a number of
authors, \cite{Ho}-\cite{EO}.

Any primordial magnetic field must be consistent with a number of
astrophysical constraints upon its strength in the early universe. Since it
provides an additional form of relativistic energy density at the epoch of
cosmological nucleosynthesis, it increases the expansion rate of the
universe; hence the neutron-proton freeze-out of weak interactions occurs at
higher temperature, with a higher value, leading to an increase in the
synthesised abundance of helium-4, \cite{T}-\cite{COST}. Therefore helium-4
observations (extrapolated to zero metallicity) lead to an upper limit on
the energy density of any cosmological magnetic field at the epoch of
nucleosynthesis. If the field is spatially homogeneous over large scales
then the COBE microwave background data provide the strongest limits on the
magnetic field strength at last scattering of the microwave background
photons, \cite{BFS}, \cite{Ba}. The strength of this limit arises from the
subtle evolution of cosmological magnetic fields in expanding universes
during the radiation era. There is a non-linear coupling between the
magnetic density and the anisotropic distortion of the expansion needed to
support it. The anisotropic pressure created by the magnetic field dominates
the evolution of the shear anisotropy and it decays far more slowly than if
the pressure is isotropic, \cite{Z}, \cite{Ba}. This shear distortion
determines the microwave background anisotropy directly. If the field is
inhomogeneous, then the limits weaken, and on small scales the fluctuations
can be dissipated. A number of discussions of the damping processes have
recently appeared, \cite{BEO}, \cite{O}, and there is a possibility that a
field of sufficient strength might have observable effects on the pattern of
Doppler peaks expected in the microwave background on small scales, \cite
{ADGR}, but the field strengths necessary may not be compatible with the
nucleosynthesis and COBE limits \cite{BFS}, \cite{Ba}.

In this paper we shall focus upon the problem of establishing a formalism
for studying the general-relativistic evolution of magnetic inhomogeneities
in an expanding universe. We do this by extending the gauge-invariant
treatment of cosmological perturbations, introduced by Ellis and Bruni for
perfect fluid cosmologies, to the case of electric and magnetic fields in a
universe which also contains a perfect fluid. This formalism identifies a
combination of gauge-invariant variables which specify the evolution of the
perturbations in an invariant manner. It offers several advantages over the
gauge-invariant approach of Bardeen, most notably by virtue of its
transparent physical interpretation.

The study of small perturbations to an homogeneous and isotropic
Friedmann-Robertson-Walker (FRW) universe is beset by subtleties of
interpretation because of gauge dependencies. The problems arise because two
different spacetimes are employed: a `smooth' spacetime, $\overline{W}$,
corresponding to the unperturbed background universe, and a `lumpy'
spacetime, $W$, that is `close' to $\overline{W}$ both dynamically and
kinematically, and represents the `perturbed' universe we live in. These two
spacetimes must be related in a way that permits the unique recovery of $%
\overline{W}$ from $W$ and vice-versa. This is impossible when the only
information provided is that the `distance' between them is small in some
suitable sense \cite{EB}. Instead, one must establish a one-to-one map, $%
\Phi :\overline{W}\rightarrow W,$ so that every point, $\overline{P,}$ in $%
\overline{W}$ has a unique image, $P,$ in $W$ and vice-versa. Establishing
such a correspondence, that is, a `gauge' (or a diffeomorphism) between the
two manifolds $\overline{W}$ and $W$, is equivalent to introducing a
fictitious smooth background model into the universe. The perturbation in
the density $\mu $ at some point $P$ of $W$ will be $\delta \mu =\mu -%
\overline{\mu }$, with $\overline{\mu }$ the density at $\overline{P}=\Phi
^{-1}(P)$ in $\overline{W}$. Clearly, there exists a relation between the
gauge choice and the value of the perturbation itself. Any variation in the
correspondence between $\overline{W}$ and $W$, which keeps the background
model fixed, is called a {\it gauge transformation}, and it is to be
distinguished from a {\it coordinate transformation} which merely relabels
events in the two spacetimes. A gauge transformation not only induces a
coordinate transformation, but also changes the point in $\overline{W}$ that
corresponds to a given point in $W$. As a result, the perturbation value
will depend on the gauge choice if the perturbed quantity is non-zero and
its spacetime coordinates will depend on the background universe \cite{B}.
Even scalars which depend on spacetime coordinates, such as the density
itself (i.e. $\mu =\mu (t)$ in $\overline{W}$), are affected by changes in
the gauge. This creates a problem, since by varying $\Phi $ one could give
the perturbation any desired value at any particular point of $W$ \cite{EB}.

There exist natural choices of gauge (e.g. fundamental fluid flow lines in
both $\overline{W}$ and $W$, hypersurfaces of simultaneity and homogeneity
in $\overline{W}$). Unfortunately, some residual freedom remains, which is
enough to create the aforementioned arbitrariness in the perturbation value.
We could deal with this gauge freedom simply by keeping track of its
consequences, but this is rather inefficient. Alternatively, we could
specify the gauge completely at the beginning and perform all calculations
in the chosen gauge. In practice however, difficulties emerge because
different gauges are best suited for different applications. A third
approach, which side-steps the gauge problem altogether, is to tackle
cosmological perturbations by means of quantities that are entirely
independent of the way the two spacetimes $\overline{W}$ and $W$ are
associated with each other.

A covariant gauge-invariant analysis of density fluctuations was introduced
by Ellis and Bruni, and initially applied to a dust-dominated
almost-Friedmann universe \cite{EB}. Their geometrical approach is simple
and physically transparent. In contrast to the standard approach, which
compares the evolution of the perturbed quantities along the worldlines of
two observers, one in $\overline{W}$ and the other in $W,$ brought together
via some specific gauge, the covariant gauge-invariant technique compares
the evolutions of neighbouring observers in the same universe, $W$. It
provides exact, fully non-linear, propagation equations for all variables,
which can be linearized about a ``variety'' of background models. This
formalism has already been extended to the study of density irregularities
within almost-FRW spacetimes with pressure \cite{EHB}, \cite{EBH}, or with a
multi-component fluid \cite{D1}, as well as to an almost-Bianchi type I
universe \cite{D2}. Here, we shall develop the covariant gauge-invariant
technique for a universe that contains a perfect fluid and large-scale
magnetic and electric fields. This provides, for the first time, a fully
general relativistic treatment of  electromagnetic perturbations for use in
cosmological investigations.

The outline of our analysis is as follows: in sections 2 and 3 the covariant
formalism is introduced. In section 4, the key gauge-invariant variables are
defined. Three of them, representing spatial changes in the energy density,
the pressure, and the expansion within an almost FRW universe, were first
introduced in \cite{EB}. The fourth gauge-invariant variable is new: it
determines variations in the magnetic field between neighbouring fundamental
observers in almost Friedmannian space-times. The medium, a single perfect
fluid of infinite conductivity, is specified in section 5, together with the
relevant equations that characterize its behaviour. In section 6, Einstein's
field equations are employed to derive the fully non-linear exact
propagation formulae for the basic kinematic quantities and for all the
gauge-invariant variables. Some aspects of the spatial geometry are briefly
discussed in section 7, and in section 8 we linearize the general equations
about a FRW universe containing a homogeneous magnetic field. Such an
assumption is a valid approximation provided that the field is ``weak'',
with an energy density much smaller than that of the isotropic perfect
fluid, which does not destroy the model's isotropy. Our results are compared
with earlier non-relativistic treatments by Ruzmaikina and Ruzmaikin \cite
{RR} and by Wasserman \cite{W}. When the pressure terms are removed from our
equations, the standard results of the Newtonian approach appear naturally.
We consider several particular applications of the formalism: the case of a
dust universe; the effect of a magnetic field on a period of de Sitter
inflation; and the growth of isocurvature perturbations.\\ 

\section{Spacetime Splitting}

%~~~~~~~~~~~~~~~~~~~~~~~~~~~~
Assume that there exists a well defined average velocity vector in the
universe. As a result, spacetime is supplied with a preferred vector field $%
u_i$ corresponding to a congruence of worldlines, known as the {\it %
fundamental fluid-flow lines}, that carry a family of special observers,
namely the {\it fundamental observers}.\footnote{%
Hereafter, an observer will always be a fundamental one unless stated
otherwise} As usual, the velocity 4-vector is normalized so that $u_i
u^i=-c^2$.

The existence of the timelike vector field $u_i$, defined by the motion of
the matter in the universe, generates a unique splitting of spacetime. To
every observer there corresponds, at any instant, a tangent spacelike
three-dimensional hypersurface $\Sigma _{\perp }$, orthogonal to $u_i$. Such
3-surfaces are the observers' instantaneous rest-spaces and they are
generally different from each other. All these 3-spaces mesh together to
create a single hyperplane, the common rest-space for all the fundamental
observers, only in a non-rotating universe.

The metric of $\Sigma _{\perp }$ is provided by a second-order symmetric
spacelike tensor, namely the observer's ``projection tensor'', defined by

\begin{equation}
h_{ij}= g_{ij}+ \frac{1}{c^2}u_i u_j ,  \label{pt}
\end{equation}
\\and satisfying

\begin{eqnarray}
h_i^{\hspace{1mm}j}&=&h_i^{\hspace{1mm}k}h_k^{\hspace{1mm}j} ,  \label{h1} \\
\nonumber \\
h_i^{\hspace{1mm}i}&=&3 .  \label{h2}
\end{eqnarray}

The projection tensor, together with the velocity 4-vector, decompose tensor
fields and tensor equations into their spatial and temporal parts. They
split the covariant derivative of $u_i$ into irreducible basic kinematic
quantities, \cite{E}, \cite{E2},

\begin{equation}
u_{i;j}=\sigma_{ij}+ \omega_{ij}+ \frac{\Theta}{3}h_{ij}- \frac{1}{c^2}\dot{%
u_i}u_j ,  \label{dui;j}
\end{equation}
\\where $\sigma_{ij}=h_{(i}^{\hspace{2mm}k}h_{j)}^{\hspace{2mm}q}u_{k;q}-
\Theta h_{ij}/3$, $\omega_{ij}=h_{[i}^{\hspace{2mm}k}h_{j]}^{\hspace{2mm}%
q}u_{k;q}$, $\Theta=u^i_{\hspace{1mm};i}$ and $\dot{u_i}=u_{i;j}u^j$ are
respectively the shear tensor, the vorticity tensor, the expansion scalar
and the 4-acceleration. The first three terms in (\ref{dui;j}) comprise the
second order tensor $v_{ij}=h_i^{\hspace{1mm}k}h_j^{\hspace{1mm}q}u_{k;q}$,
which in turn determines the relative spatial velocity between neighbouring
worldlines. The expansion scalar is used to introduce a representative
length scale ($S$) along the observer's worldline. In particular, we define

\begin{equation}
\frac{\dot{S}}{S}=\frac{\Theta}{3} ,  \label{HubL}
\end{equation}
\\for the volume expansion.

The Weyl tensor, $C_{ijkq}$, is decomposed into a pair of symmetric
traceless and completely spacelike second-order tensors, $E_{ij}$ and $H_{ij}
$, respectively known as the ``electric'' and the ``magnetic'' parts of $%
C_{ijkq}$ \cite{H}, \cite{E2},

\begin{eqnarray}
E_{ij}&=& \frac{1}{c^2}C_{ikjq}u^k u^q ,  \label{el} \\
\nonumber \\
H_{ij}&=& \frac{1}{2c^2}\eta_{ip}^{\hspace{2mm}kq}C_{kqjs}u^p u^s .
\label{mag}
\end{eqnarray}
\\where, $\eta_{ijkq}$ is the covariant permutation tensor of the spacetime.

\section{The Electromagnetic Field}

%~~~~~~~~~~~~~~~~~~~~~~~~~~~~~~~~~~

\subsection{The Electromagnetic Field Tensor}

%~~~~~~~~~~~~~~~~~~~~~~~~~~~~~~~~~~~~~~~~~~~~
The electromagnetic field is represented by the second-order antisymmetric
Maxwell tensor $F_{ij}$, satisfying

\begin{equation}
F_{[ij;k]}= F_{ij;k}+F_{jk;i}+F_{ki;j}=0 .  \label{Ften}
\end{equation}

The components of $F_{ij}$ are derived from a potential, $V_i$, according to

\begin{equation}
F_{ij}=V_{j;i}-V_{i;j}=\frac{\partial V_j}{\partial x^i}-\frac{\partial V_i}{%
\partial x^j},  \label{emf}
\end{equation}
\\ where $V_i=\left( V_\mu ,-\Psi /c\right) $ are the covariant components
of the 4-potential relative to a freely-falling frame. The quantities $V_\mu 
$ and $\Psi $ are the vector and the scalar potentials respectively.\\ 

\subsection{The Electromagnetic Field Components}

%~~~~~~~~~~~~~~~~~~~~~~~~~~~~~~~~~~~~~~~~~~~~~~~~
As seen by an observer moving with 4-velocity $u_i$, the electromagnetic
field tensor decomposes into an electric ($E_i$) and a magnetic ($H_i$)
part, defined by \cite{E2},

\begin{equation}
E_i=F_{ij}u^j=-F_{ij}u^i,  \label{ef}
\end{equation}
\\ and 
\begin{equation}
H_i=\frac 1{2c}\eta _{ijkq}u^jF^{kq},  \label{mf}
\end{equation}
\\ respectively. From definitions (\ref{ef}), (\ref{mf}) and the total
skewness of $F_{ij}$ and $\eta _{ijkq}$, we obtain

\begin{eqnarray}
E_i u^i&=&0 ,  \label{P1} \\
\nonumber \\
H_i u^i&=&0 ,  \label{P2}
\end{eqnarray}
\\to ensure that both $E_i$ and $H_i$ are 3-vectors in the observer's rest
space.

The projection of $F_{ij}$ onto the 3-surface $\Sigma_{\perp}$, the
observer's instantaneous rest space, provides the decomposition

\begin{equation}
F_{ij}= \frac{1}{c^2}\left(u_i E_j-E_i u_j\right)- \frac{1}{c}\eta_{ijkq}u^k
H^q ,  \label{demf}
\end{equation}
\\and shows how $E_i$ and $H_i$ completely determine $F_{ij}$.

\subsection{Maxwell's Equations}

%~~~~~~~~~~~~~~~~~~~~~~~~~~~~~~~
The first of Maxwell's equations has the form

\begin{equation}
F^{ij}_{\hspace{2mm};j}=\frac{1}{c}J^i ,  \label{Max1}
\end{equation}
\\where $J_i$ is the 4-current which generates the electromagnetic field and
obeys the conservation law $J^i_{\hspace{1mm};i}=0$. The second equation,

\begin{equation}
F_{[ij;k]}=0 ,  \label{Max2}
\end{equation}
\\is a direct consequence of the existence of the 4-potential.

As measured by a fundamental observer equipped with the projection tensor $%
h_{ij}$, equations (\ref{Max1}) and (\ref{Max2}) decompose into temporal and
spatial parts as \cite{E2}

\begin{eqnarray}
E_{\hspace{1mm};i}^i+2\omega ^iH_i-\frac 1{c^2}E^i\dot u_i &=&\epsilon c,
\label{M1} \\
&&  \nonumber \\
\left( \sigma _{\hspace{1mm}j}^i+\omega _{\hspace{1mm}j}^i-\frac{2\Theta }%
3h_{\hspace{1mm}j}^i\right) E^j+\frac 1c\eta ^{ijkq}u_j\dot u_kH_q-c\eta
^{ijkq}u_jH_{k;q}-ch_{\hspace{1mm}j}^iJ^j &=&h_{\hspace{1mm}j}^i\dot E^j,
\label{M2} \\
&&  \nonumber \\
H_{\hspace{1mm};i}^i-\frac 2{c^2}\omega ^iE_i-\frac 1{c^2}H^i\dot u_i &=&0,
\label{M3} \\
&&  \nonumber \\
\left( \sigma _{\hspace{1mm}j}^i+\omega _{\hspace{1mm}j}^i-\frac{2\Theta }%
3h_{\hspace{1mm}j}^i\right) H^j-\frac 1{c^3}\eta ^{ijkq}u_j\dot u_kE_q+\frac
1c\eta ^{ijkq}u_jE_{k;q} &=&h_{\hspace{1mm}j}^i\dot H^j,  \label{M4}
\end{eqnarray}
\\ where the vorticity vector ($\omega _i$), and the charge density ($%
\epsilon =-J^iu_i/c^2$) are new variables. Note that $\omega _{ij}$ and $%
\omega _i$ are related by the formula

\begin{equation}
\omega_{ij}=\frac{1}{c}\eta_{ijkq}\omega^k u^q ,  \label{vt}
\end{equation}
\\which ensures that $\omega_i\omega^{ij}=0$.\\

\subsection{The Energy-Momentum Tensor of the Electromagnetic Field}

%~~~~~~~~~~~~~~~~~~~~~~~~~~~~~~~~~~~~~~~~~~~~~~~~~~~~~~~~~~~~~~~~~~~
The electromagnetic energy-momentum tensor ($T^{ij}_{em}$) is a symmetric
and trace-free second-order tensor of the following general form

\begin{equation}
T^{ij}_{em}= F^{ki}F_k^{\hspace{1mm}j}- \frac{1}{4}g^{ij}F_{kq}F^{kq} .
\label{Tem}
\end{equation}

Relative to a fundamental observers, $T_{em}^{ij}$ splits as \cite{E2},

\begin{equation}
T_{em}^{ij}=\frac 1{2c^2}\left( \frac{E^2}{c^2}+H^2\right) u^iu^j+\frac
16\left( \frac{E^2}{c^2}+H^2\right) h^{ij}+\frac 2{c^3}u^{(i}\eta
^{j)kqs}u_kE_qH_s+M^{ij},  \label{Tem1}
\end{equation}
\\ where $E^2=E_iE^i$ and $H^2=H_iH^i$ are the magnitudes of the field's
electric and the magnetic components respectively; $M_{ij}$ is a traceless
and completely spacelike symmetric tensor given by

\begin{equation}
M^{ij}= \frac{1}{3}\left(\frac{E^2}{c^2}+H^2\right)h^{ij}- \frac{1}{c^2}E^i
E^j-H^i H^j .  \label{Mten1}
\end{equation}

The expression (\ref{Tem1}) allows a direct comparison to be made between $%
T_{em}^{ij}$ and the energy-momentum tensor, $\mbox{}^{*}T_{ij},$ of a
general imperfect fluid possessing viscous and heat conduction contributions:

\begin{equation}
\mbox{}^* T^{ij}=\mbox{}^* \mu u^i u^j+ \mbox{}^* ph^{ij}+ \frac{2}{c^2}%
\mbox{}^* q^{(i}u^{j)}+ \mbox{}^* \pi^{ij} .  \label{Tim}
\end{equation}
\\The resulting correspondence, which provides a ``fluid description'' for
the electromagnetic field, is

\begin{eqnarray}
\mu _{em} &=&\frac 1{2c^2}\left( \frac{E^2}{c^2}+H^2\right) ,  \label{muem}
\\
&&  \nonumber \\
p_{em} &=&\frac 16\left( \frac{E^2}{c^2}+H^2\right) ,  \label{pem} \\
&&  \nonumber \\
q_{em}^i &=&\frac 1c\eta ^{ijkq}u_jE_kH_q,  \label{qem} \\
&&  \nonumber \\
\pi _{em}^{ij} &=&M^{ij},  \label{piem}
\end{eqnarray}
\\ and leads to the familiar equation of state of radiation $p_\gamma =\mu
_\gamma c^2/3$. The last expression suggests that $M_{ij}$ contributes an 
anisotropic electromagnetic pressure.\\ 

\section{The Key Gauge-Invariant Variables}

%~~~~~~~~~~~~~~~~~~~~~~~~~~~~~~~~~~~~~~~~~~

\subsection{ The ${\bf D_i}$, ${\bf Y_i}$ and ${\bf {\cal Z}_i}$ Spatial
Gradients}

%~~~~~~~~~~~~~~~~~
In their covariant and gauge-invariant study of cosmological perturbations
Ellis and Bruni \cite{EB} describe spatial variations in the energy density (%
$\mu c^2$), the pressure ($p$) and the expansion ($\Theta $) by projecting
their gradients onto the instantaneous rest space of an observer comoving
with the expanding fluid. Assuming that the unperturbed background universe
is represented by a FRW spacetime, they consider the following basic
variables

\begin{eqnarray}
X_i &=&\kappa h_i^{\hspace{1mm}j}\mu _{;j}c^2=\kappa \mbox{}^{(3)}\nabla
_i\mu c^2,  \label{X} \\
&&  \nonumber \\
Y_i &=&\kappa h_i^{\hspace{1mm}j}p_{;j}=\kappa \mbox{}^{(3)}\nabla _ip,
\label{Y} \\
&&  \nonumber \\
Z_i &=&h_i^{\hspace{1mm}j}\Theta _{;j}=\mbox{}^{(3)}\nabla _i\Theta ,
\label{Z}
\end{eqnarray}
\\ where $\kappa =8\pi G/c^4$ is the Einstein gravitational constant. All
three vectors $X_i$, $Y_i$ and $Z_i$ vanish in a FRW model, thus ensuring
their gauge-invariance.\footnote{%
The simplest gauge-invariant quantities are scalars that are constant in the
background, or tensors which are zero there. The only other possibility is a
tensor written as a linear combination of products of Kronecker deltas with
constant coefficients (Stewart and Walker lemma \cite{SW}).} Ellis and Bruni
then define the following ``comoving'' variables

\begin{equation}
D_i=\frac{SX_i}{\kappa \mu c^2},  \label{D}
\end{equation}
\\ namely the comoving fractional orthogonal spatial gradient of the energy
density, and

\begin{equation}
{\cal Z}_i=SZ_i,  \label{cZ}
\end{equation}
\\ which is the comoving orthogonal spatial gradient of the expansion.
Besides being invariant under gauge transformations, $D_i$ and ${\cal Z}_i$
also describe the spatial variations of $\mu c^2$ and $\Theta $
respectively, within a perturbed FRW universe \cite{EB}.\\ 

\subsection{A Gauge-Invariant Variable for the Magnetic Field}

%~~~~~~~~~~~~~~~~~~~~~~~~~~~~~~~~~~~~~~~~~~~~~~~~~~~~~~~~~~~~~
The assumption of a large-scale primordial magnetic field requires that the
fictitious background universe must include some degree of anisotropy and a
spatially homogeneous non-rotating Bianchi-I spacetime provides the simplest
example of such a universe. However, later we shall primarily consider the
case of an unperturbed  FRW spacetime. This is an acceptable simplification
provided that the field is too weak to affect the model's isotropy and
allows us compare our approach with the existing (Newtonian) results
directly.

Although Ellis and Bruni restricted their study to almost-FRW geometries,
the gauge-invariance of their key variables also holds in any spatially
homogeneous, non-rotating background spacetime, such as a Bianchi-I
universe. Furthermore, the same variables can describe spatial gradients in $%
\mu c^2$, $p$ and $\Theta $ within an almost Bianchi-I model, as it has been
argued by Dunsby \cite{D2}.

Let us consider a spatially homogeneous universe permeated by a large-scale
magnetic field (i.e. $H_i=H_i(t)$) with no accompanying electric field (i.e. 
$E_i\equiv 0$). Following Ellis and Bruni \cite{EB}, we define the
second-order tensor

\begin{equation}
{\cal H}_{ij}=\kappa h_i^{\hspace{1mm}k}h_j^{\hspace{1mm}q}H_{k;q}=\kappa %
\mbox{}^{(3)}\nabla _jH_i,  \label{Hten}
\end{equation}
\\ namely the orthogonal spatial gradient of the magnetic field. When the
vorticity is zero, ${\cal H}_{ij}$ vanishes (see Appendix A.1), and so
provides a new gauge-invariant variable. Moreover, the quantity

\begin{equation}
{\cal M}_{ij}=S{\cal H}_{ij},  \label{MHten}
\end{equation}
\\ namely the comoving orthogonal spatial gradient of the magnetic field,
describes the spatial variations of $H_i$ between neighbouring world lines
in a perturbed FRW universe (see Appendix A.2). Following Dunsby \cite{D2},
we can extent this result to a nearly Bianchi-I spacetime if required.

The quantity ${\cal H}_{ij}$ is a completely spacelike and traceless tensor

\begin{eqnarray}
u^i{\cal H}_{ij}&=&{\cal H}_{ij}u^j=0 ,  \label{cH1} \\
\nonumber \\
{\cal H}^i_{\hspace{1mm}i}&=&0 ,  \label{cH3}
\end{eqnarray}
\\where the last result is derived from (\ref{M3}) under the assumption of a
zero electric field.

\section{Specifying the Medium}

%~~~~~~~~~~~~~~~~~~~~~~~~~~~~~~

\subsection{The Case of Infinite Conductivity}

%~~~~~~~~~~~~~~~~~~~~~~~~~~~~~~~~~~~~~~~~~~~~~
In order to specify the material content of the universe, we consider the
covariant form of Ohm's law \cite{J},

\begin{equation}
J_i+\frac 1{c^2}J^ju_ju_i=\frac 1c\overline{\sigma }E_i,  \label{cOhm}
\end{equation}
\\ where $u_i$ is the fluid velocity and $\overline{\sigma }$ represents the
conductivity of the medium. Projecting onto the instantaneous rest space of
a fundamental observer, we obtain

\begin{equation}
h_i^{\hspace{1mm}j}J_j=\frac{1}{c}\overline{\sigma}E_i .  \label{Ohm}
\end{equation}

Thus, non-zero spatial current densities (i.e. $h_i^{\hspace{1mm}j}J_j\neq 0$%
), are compatible with a vanishing electric field as long as the
conductivity of the medium is infinite (i.e. $\overline{\sigma}%
\rightarrow\infty$). Under this assumption, formulae (\ref{M1})-(\ref{M4})
become

\begin{eqnarray}
2\omega^i H_i&=&\epsilon c ,  \label{Mi1} \\
\nonumber \\
\eta^{ijkq}u_j\left(\dot{u}_k H_q-c^2 H_{k;q}\right)&=& c^2 h^i_{\hspace{1mm}%
j}J^j ,  \label{Mi2} \\
\nonumber \\
H^i_{\hspace{1mm};i}-\frac{1}{c^2}H^i \dot{u}_i&=&0 ,  \label{Mi3} \\
\nonumber \\
\left(\sigma^i_{\hspace{1mm}j}+\omega^i_{\hspace{1mm}j} -\frac{2\Theta}{3}%
h^i_{\hspace{1mm}j}\right)H^j&=& h^i_{\hspace{1mm}j}\dot{H}^j .  \label{Mi4}
\end{eqnarray}

Equation (\ref{Mi3}) can be rearranged as

\begin{equation}
h_i^{\hspace{1mm}j}H_{\hspace{1mm};j}^i=\mbox{}^{(3)}\nabla _iH^i=0,
\label{Mi12}
\end{equation}
\\ to provide the familiar 'vanishing 3-divergence' of the magnetic field.
Some useful expressions can be obtained from (\ref{Mi4}); in particular,

\begin{equation}
h_{\hspace{1mm}j}^i\left( S^2H^j\right) ^{\cdot }=\left( \sigma _{%
\hspace{1mm}j}^i+\omega _{\hspace{1mm}j}^i\right) S^2H^j,  \label{Mi41}
\end{equation}
\\ is the covariant analogue of the `induction' equation and verifies that
within an exactly FRW universe the magnetic field decays adiabatically as
the inverse square of the scale factor. An additional formula, which will be
used later to simplify the energy-density conservation law, is obtained
after contracting (\ref{Mi4}) with $H_i$

\begin{equation}
\sigma_{ij}H^i H^j=\frac{2\Theta H^2}{3}+ \frac{\left(H^2\right)^{\cdot}}{2}
.  \label{Mi42}
\end{equation}

The introduction of a perfectly conducting medium, as opposed to the
assumption of a pure magnetic field with no electric field and zero spatial
currents, is important in this approach. As we shall see, it allows
perturbations in the energy density of the fluid to be coupled with magnetic
irregularities in a straightforward and natural way (see further discussion
in Appendix B).

\subsection{Magnetized Perfect Fluid with Infinite Conductivity}

%~~~~~~~~~~~~~~~~~~~~~~~~~~~~~~~~~~~~~~~~~~~~~~~~~~~~~~~~~~~~~~~~~~~~
In a medium with infinite conductivity (i.e. $E_i=0$), the electromagnetic
energy-momentum tensor (see (\ref{Tem1})) takes the form

\begin{equation}
T^{ij}_{em}=\frac{H^2}{2c^2}u^i u^j+ \frac{H^2}{6}h^{ij}+ M^{ij} ,
\label{Tem2}
\end{equation}
\\where

\begin{equation}
M^{ij}=\frac{H^2}{3}h^{ij}-H^iH^j .  \label{Mten2}
\end{equation}

The energy-momentum tensor of a single perfect fluid (i.e. zero energy-flux
and no anisotropic stresses) is

\begin{equation}
T^{ij}_m=\mu u^i u^j+ph^{ij} .  \label{Tm}
\end{equation}

Thus, the energy-momentum tensor that describes a single magnetized perfect
fluid of infinite conductivity has the form

\begin{equation}
T^{ij}=\left(\mu+\frac{H^2}{2c^2}\right)u^i u^j+ \left(p+\frac{H^2}{6}%
\right)h^{ij}+ M^{ij} ,  \label{Tten}
\end{equation}
\\with trace $T=3p-\mu c^2$.

A comparison between (\ref{Tten}) and (\ref{Tim}), the stress tensor of an
imperfect fluid, shows that a single magnetized perfect fluid of infinite
conductivity can be represented as an imperfect fluid with the following
properties

\begin{eqnarray}
\mbox{}^* \mu&=&\mu+\frac{H^2}{2c^2} ,  \label{cor1} \\
\nonumber \\
\mbox{}^* p&=&p+\frac{H^2}{6} ,  \label{cor2} \\
\nonumber \\
\mbox{}^*q&=&0 ,  \label{cor4} \\
\nonumber \\
\mbox{}^* \pi^{ij}&=&M^{ij}=\frac{H^2}{3}h^{ij}-H^iH^j .  \label{cor3}
\end{eqnarray}

We can now derive the conservation laws that characterize the medium. We
start by setting the divergence of (\ref{Tten}) equal to zero. The resulting
formula is then split into a temporal and a spatial part, respectively
expressing the conservation of the energy and of the momentum densities.
More precisely, contracting with the observer's 4-velocity and using (\ref
{Mi42}) we find

\begin{equation}
\dot{\mu}c^2+\left(\mu c^2+p\right)\Theta=0 ,  \label{edc}
\end{equation}
\\which coincides with the energy-density conservation law for a single
non-magnetized perfect fluid. The conservation of the momentum-density is
obtained after projecting (\ref{Tten}) onto the observer's instantaneous
rest space,

\begin{equation}
\left(\mu+\frac{p}{c^2}+\frac{H^2}{c^2}\right)\dot{u}^i+ h^{ij}\left(p+\frac{%
H^2}{2}\right)_{;j}- h^i_{\hspace{1mm}j}H^j_{\hspace{1mm};k}H^k- H^i H^j_{%
\hspace{1mm};j}=0 ,  \label{mdc}
\end{equation}
\\and can be rewritten, in order to involve the spatial gradients $Y_i$ and $%
{\cal H}_{ij}$, as

\begin{equation}
\kappa\left(\mu+\frac{p}{c^2}+\frac{2H^2}{3c^2}\right)\dot{u}_i+ Y_i- 2{\cal %
H}_{[ij]}H^j+ \frac{\kappa}{c^2}\dot{u}^j M_{ji}=0 .  \label{mdc1}
\end{equation}

Equation (\ref{mdc}) (or (\ref{mdc1})) naturally connects spatial
fluctuations in the magnetic field, with the acceleration and gradients in
the pressure, and subsequently with spatial variations in the energy density
(see also Appendix B).\\

\section{The General Propagation Equations}

%~~~~~~~~~~~~~~~~~~~~~~~~~~~~~~~~~~~~~~~~~~

\subsection{The Field Equations}

%~~~~~~~~~~~~~~~~~~~~~~~~~~~~~~~
The general form of Einstein's field equations is

\begin{equation}
R^{ij}=\kappa T^{ij}-\frac{\kappa T}{2}g^{ij}+\Lambda g^{ij} ,  \label{fes}
\end{equation}
\\where $R_{ij}=R_{i\hspace{1mm}jk}^{\hspace{1mm}k}$ is the Ricci tensor, $%
g_{ij}$ is the spacetime metric and $\Lambda$ is the cosmological constant.
The Ricci scalar ($R=R^i_{\hspace{1mm}i}$) and the trace ($T$) of the
energy-momentum tensor, that describes the material content, are related by

\begin{equation}
R=4\Lambda-\kappa T .  \label{traces}
\end{equation}

Using the observer's projection tensor, the Ricci tensor splits as

\begin{equation}
R^{ij}=h_{\hspace{1mm}k}^ih_{\hspace{1mm}q}^jR^{kq}-\frac 1{c^2}\left( h_{%
\hspace{1mm}q}^iR^{qk}u_ku^j+u^iu_kR^{kq}h_q^{\hspace{1mm}j}\right) +\frac
1{c^4}R^{kq}u_ku_qu^iu^j,  \label{Rd}
\end{equation}
\\ which shows that the decomposition is entirely determined by the sums $h_{%
\hspace{1mm}k}^ih_{\hspace{1mm}q}^jR^{kq}$, $h_{\hspace{1mm}j}^iR^{jk}u_k$
and $R^{ij}u_iu_j$.

When dealing with a single perfect fluid of infinite conductivity, permeated
by a large-scale magnetic field, $T_{ij}$ is given by (\ref{Tten}) and
consequently we find

\begin{eqnarray}
h^i_{\hspace{1mm}k}h^j_{\hspace{1mm}q}R^{kq}&=& \left(\frac{\kappa}{2}%
\left(\mu c^2-p+\frac{H^2}{3}\right) +\Lambda\right)h^{ij}+ \kappa M^{ij} ,
\label{Rdsp} \\
\nonumber \\
h^i_{\hspace{1mm}j}R^{jk}u_k&=&0 ,  \label{Rdst} \\
\nonumber \\
R^{ij}u_i u_j&=& \frac{\kappa c^2}{2}\left(\mu c^2+3p+H^2\right)- \Lambda
c^2 .  \label{Rdtem}
\end{eqnarray}

Thus, the Ricci tensor associated with our cosmological model, decomposes
into purely spatial and temporal parts only,

\begin{equation}
R^{ij}=\left(\frac{\kappa}{2}\left(\mu c^2-p+\frac{H^2}{3}\right)
+\Lambda\right)h^{ij}+ \kappa M^{ij}+ \left(\frac{\kappa}{2}\left(\mu+\frac{%
3p}{c^2}+\frac{H^2}{c^2}\right)- \frac{\Lambda}{c^2}\right)u^i u^j ,
\label{Rd1}
\end{equation}
\\while the Ricci scalar becomes

\begin{equation}
R=\kappa\left(\mu c^2-3p\right)+4\Lambda .  \label{Rsc}
\end{equation}

These equations allow us to derive the formulae that determine the
time-evolution of the expansion, the shear and the vorticity.\\

\subsection{Propagation of the Kinematic Quantities}

%~~~~~~~~~~~~~~~~~~~~~~~~~~~~~~~~~~~~~~~~~~~~~~~~~~~
The evolution of basic kinematic variables such as the expansion scalar ($%
\Theta$), the shear tensor ($\sigma_{ij}$) and the vorticity vector ($%
\omega_i$), along the world line of a fundamental observer, is governed by
their respective propagation equations. These are derived from the following
formula, describing the evolution of the relative velocity between
neighbouring fundamental fluid-flow lines

\begin{equation}
h_i^{\hspace{1mm}k}h_j^{\hspace{1mm}q}\dot v_{kq}-\frac 1{c^2}\dot u_i\dot
u_j-\mbox{}^{(3)}\nabla _j\dot u_i+v_{ik}v_{\hspace{1mm}%
j}^k+R_{ikjq}u^ku^q=0.  \label{rvpr}
\end{equation}
Contracting (\ref{rvpr}), recalling that $v_{ij}=\sigma _{ij}+\omega
_{ij}+\Theta h_{ij}/3,$ and using (\ref{Rdtem}), we find that under the
influence of a cosmological magnetic field Raychaudhuri's relation becomes

\begin{equation}
\dot{\Theta}-A+ 2\left(\sigma^2-\omega^2\right) +\frac{\Theta^2}{3}+ \frac{%
\kappa c^2}{2}\left(\mu c^2+3p+H^2\right)- \Lambda c^2=0 ,  \label{Ray}
\end{equation}
\\where $A=\dot{u}^i_{\hspace{1mm};i}$, $\sigma^2=\sigma^{ij}\sigma_{ij}/2$
and $\omega^2=\omega^{ij}\omega_{ij}/2$ by definition.

The shear propagation formula is obtained by taking the symmetric,
trace-free, part of (\ref{rvpr}). In order to derive the final expression,
we employ (\ref{Ray}), the definition of the Weyl tensor along with
equations ({\ref{Rdsp}), (\ref{Rdtem}) and (\ref{Rsc}). The outcome is }

\begin{eqnarray}
h^{ik}h^{jq}\left(\dot{\sigma}_{kq}-\dot{u}_{(k;q)}\right)+ \frac{1}{3}%
\left(A-2\sigma^2-\omega^2\right)h^{ij}- \frac{\kappa c^2}{2}M^{ij}- \frac{1%
}{c^2}\dot{u}^i\dot{u}^j+ \sigma^{ik}\sigma_k^{\hspace{1mm}j}+&\mbox{}& 
\nonumber \\
\frac{2\Theta}{3}\sigma^{ij}+ \omega^i\omega^j+E^{ij}&=&0 ,  \label{shpr}
\end{eqnarray}
\\where $E_{ij}$ is the electric part of the Weyl tensor and $M_{ij}$ is
given by (\ref{Mten2}). In deriving (\ref{shpr}) we have used the result $%
\omega_{ik}\omega^k_{\hspace{1mm}j}=\omega_i\omega_j-\omega^2 h_{ij}$, which
relates $\omega^2$ to the vorticity tensor and vector. Comparing (\ref{shpr}%
) with the corresponding formula in the case of a non-magnetized imperfect
fluid (see \cite{E2}), we notice again that the distortion generated by the
field is analogous to that induced by anisotropic stresses. The proportional
to the projection tensor term, simply subtracts off the trace.

Finally, the skew part of (\ref{rvpr}), provides the equation governing the
rate of change of the vorticity vector

\begin{equation}
h^i_{\hspace{1mm}j}\left(S^2\omega^j\right)^{\cdot}= \frac{S^2}{2c}%
\eta^{ijkq}u_j\dot{u}_{k;q}+ S^2\omega^j\sigma_j^{\hspace{1mm}i} ,
\label{vorpr}
\end{equation}
\\where, $\omega^i=\eta^{ijkq}u_j\omega_{kq}/2c$. Notice that the magnetic
effects are felt indirectly through the shear and the acceleration.\\

\subsection{Propagation Equations for the Key Variables}

%~~~~~~~~~~~~~~~~~~~~~~~~~~~~~~~~~~~~~~~~~~~~~~~~~~~~~~~
From definitions (\ref{X})-(\ref{cZ}) we can derive the propagation formulae
for the various spatial gradients. The most important equation determines
the evolution of the comoving fractional orthogonal spatial energy-density
gradient ($D_i$)

\begin{eqnarray}
h_i^{\hspace{1mm}j}\dot D_j &=&\frac{p\Theta }{\mu c^2}D_i-D_j\left( \sigma
_{\hspace{1mm}i}^j+\omega _{\hspace{1mm}i}^j\right) -\left( 1+\frac p{\mu
c^2}\right) {\cal Z}_i-\frac{2\Theta }{\kappa \mu c^2}{\cal M}_{[ij]}H^j+ 
\nonumber \\
&&\mbox{}\frac{2S\Theta H^2}{3\mu c^4}\dot u_i+\frac{S\Theta }{\mu c^4}\dot
u^jM_{ji}.  \label{Dprop}
\end{eqnarray}
\\ It is obtained by taking the time derivative of (\ref{D}) and then using
the conservation laws (\ref{edc}) and (\ref{mdc1}). Notice that the field
gradients are already connected with those in the energy density, unlike the
Newtonian approach, where the coupling occurs at the next level. This
coupling arises from the presence of the quantity $\dot u_i$, which will
then bring $D_i$ and ${\cal M}_{ij}$ together via the momentum-density
conservation law.

The vector ${\cal Z}_i$, the comoving orthogonal spatial gradient of the
expansion, has a rate of change determined by the following propagation
formula,

\begin{eqnarray}
h_i^{\hspace{1mm}j}\dot{{\cal Z}}_j&=& -\frac{2\Theta}{3}{\cal Z}_i- {\cal Z}%
_j\left(\sigma^j_{\hspace{1mm}i}+\omega^j_{\hspace{1mm}i}\right)- \kappa\mu
c^4 \left(\frac{1}{2}D_i+\frac{1}{\kappa\mu c^2}{\cal M}_{ji}H^j\right)- 3c^2%
{\cal M}_{[ij]}H^j+  \nonumber \\
&\mbox{}&S{\cal R}\dot{u}_i+ \frac{3\kappa S}{2}\dot{u}^j M_{ji}+ SA_i- 2S%
\mbox{}^{(3)}\nabla_i \left(\sigma^2-\omega^2\right) ,  \label{cZprop}
\end{eqnarray}
\\where $A_i =h_i^{\hspace{1mm}j}A_{;j}=\mbox{}^{(3)}\nabla_i A$, and

\begin{equation}
{\cal R}=\left( \kappa \left( \mu c^2+\frac{H^2}2\right) -\frac \Theta
{3c^2}+\frac{\sigma ^2}{c^2}-\frac{\omega ^2}{c^2}+\Lambda \right) +\frac
A{c^2}-\frac 3{c^2}\left( \sigma ^2-\omega ^2\right) ,  \label{R}
\end{equation}
\\ with the quantity in brackets representing, as we shall see in section 7,
the 3-Ricci scalar of the observer's instantaneous rest space. To derive (%
\ref{cZprop}) we start with the time derivative of (\ref{cZ}), and then use (%
\ref{mdc1}), together with (\ref{Ray}).

The third key variable is the comoving orthogonal spatial gradient of the
magnetic field (${\cal M}_{ij}$). In order to derive the propagation formula
for ${\cal M}_{ij}$, we first split the covariant derivative of the field
with respect to irreducible kinematic quantities, employing the observer's
projection tensor and equation (\ref{Mi4}) to obtain,

\begin{equation}
H_{i;j}=\frac 1\kappa {\cal H}_{ij}-\frac 2{c^2}H_k\sigma _{\hspace{1mm}%
[i}^ku_{j]}+\frac 2{c^2}H_k\omega _{\hspace{1mm}(i}^ku_{j)}+\frac \Theta
{3c^2}\left( 2H_iu_j+u_iH_j\right) -\frac 1{c^4}\dot u^kH_ku_iu_j.
\label{dHi;j}
\end{equation}
\\ The time evolution of the spatial gradient of the magnetic field can now
be obtained by applying the Ricci identity to the field vector, in
connection with (\ref{dHi;j}). The outcome,

\begin{eqnarray}
S^{-2}h_i^{\hspace{1mm}k}h_j^{\hspace{1mm}q}\left( S^2{\cal M}_{kq}\right)
^{\cdot } &=&-{\cal M}_{ik}\left( \sigma _{\hspace{1mm}j}^k+\omega _{%
\hspace{1mm}j}^k\right) +\left( \sigma _i^{\hspace{1mm}k}+\omega _i^{%
\hspace{1mm}k}\right) {\cal M}_{kj}-\frac{2\kappa }3H_i{\cal Z}_j-  \nonumber
\\
&&\mbox{}\frac{2\kappa S}{c^2}H_k\omega _{\hspace{1mm}(i}^k\dot u_{j)}+\frac{%
2\kappa S}{c^2}H_k\sigma _{\hspace{1mm}[i}^k\dot u_{j]}+  \nonumber \\
&&\mbox{}\kappa SH^k\mbox{}^{(3)}\nabla _j\left( \sigma _{ik}+\omega
_{ik}\right) -\frac{\kappa \Theta S}{3c^2}\left( 2H_i\dot u_j+\dot
u_iH_j\right) +  \nonumber \\
&&\mbox{}\frac{\kappa S}{c^2}\dot u^kH_k\left( \sigma _{ij}+\omega
_{ij}+\frac \Theta 3h_{ij}\right) -\kappa Sh_i^{\hspace{1mm}k}R_{kqjs}H^qu^s,
\label{Mprop}
\end{eqnarray}
\\ shows how the Riemann curvature tensor $R_{ijkq}$ acts as an additional
source for magnetic inhomogeneities. Notice that so far there has been no
formula determining the rate of change of spatial gradients in the pressure.
When an equation of state for the non-magnetic fluid is introduced, the
propagation of $Y_i$, along the observer's worldline, can be obtained from (%
\ref{Dprop}).

Dunsby~\cite{D1} applied the covariant and gauge-invariant technique to the
study of cosmological perturbations in an almost FRW universe filled with an
imperfect non-magnetized fluid. As mentioned earlier, our stress can be
written as corresponding to an imperfect fluid with the special properties
given by (\ref{cor1})-(\ref{cor3}). When Dunsby's fully non-linear
propagation formulae for $D_i$ and ${\cal Z}_i$ are applied to this
particular imperfect fluid, we recover equations (\ref{Dprop}) and (\ref
{cZprop}). Of course, there is no expression equivalent to (\ref{Mprop}) in
Dunsby's analysis.\\ 

\section{Aspects of the Spatial Geometry}

%~~~~~~~~~~~~~~~~~~~~~~~~~~~~~~~~~~~~~~~~
The three-dimensional Riemann tensor ($\mbox{}^{(3)}R_{ijkq}$), that
determines the curvature of a fundamental observer's instantaneous rest
space, $\Sigma _{\perp }$, is defined via the commutator for the 3-D
gradients of any spacelike vector. In particular, provided that $v_iu^i=0$,
we have \cite{EBH},

\begin{equation}
\mbox{}^{(3)}\nabla _i\mbox{}^{(3)}\nabla _jv_k-\mbox{}^{(3)}\nabla _j\mbox{}%
^{(3)}\nabla _iv_k=-\frac 2{c^2}\omega _{ij}h_k^{\hspace{1mm}q}\dot v_q+%
\mbox{}^{(3)}R_{qkji}v^q,  \label{3Rid}
\end{equation}
\\ where, by definition

\begin{equation}
\mbox{}^{(3)}R_{ijkq}=h_i^{\hspace{1mm}p}h_j^{\hspace{1mm}s}h_k^{\hspace{1mm}%
r}h_q^{\hspace{1mm}t}R_{psrt}-\frac 1{c^2}v_{ik}v_{jq}+\frac
1{c^2}v_{iq}v_{jk},  \label{3cten}
\end{equation}
\\ with $v_{ij}=\mbox{}^{(3)}\nabla _ju_i=\sigma _{ij}+\omega _{ij}+\Theta
h_{ij}/3$. The successive contractions of (\ref{3cten}) provide the 3-Ricci
tensor ($\mbox{}^{(3)}R_{ij}$) and the 3-Ricci scalar ($\mbox{}^{(3)}R$) of $%
\Sigma _{\perp }$, by 

\begin{equation}
\mbox{}^{(3)}R_{ij}= h_i^{\hspace{1mm}k}h_j^{\hspace{1mm}q}R_{kq}+ \frac{1}{%
c^2}R_{ikjq}u^k u^q+ \frac{1}{c^2}v_{ik}v^k_{\hspace{1mm}j}- \frac{\Theta}{%
c^2}v_{ij} ,  \label{3Rten}
\end{equation}
\\and

\begin{equation}
\mbox{}^{(3)}R=R+\frac 2{c^2}R_{ij}u^iu^j-\frac{2\Theta ^2}{3c^2}+\frac{%
2\sigma ^2}{c^2}-\frac{2\omega ^2}{c^2}.  \label{3Rsc}
\end{equation}
Notice the vorticity term that appears in equation (\ref{3Rid}). In a
rotating spacetime, this term prevents the commutation between the
3-gradients of scalars. In particular, the formula

\begin{equation}
\mbox{}^{(3)}\nabla_i\mbox{}^{(3)}\nabla_j f- \mbox{}^{(3)}\nabla_j\mbox{}%
^{(3)}\nabla_i f= -\frac{2}{c^2}\omega_{ij}\dot{f} ,  \label{3sc}
\end{equation}
\\holds for any scalar quantity $f$. This non-commutativity, reflects the
fact that the fluid velocity does not consist a hypersurface orthogonal
vector field if $\omega_{ij}\neq0$ (see Appendix in \cite{EBH} for more
details).

Equations (\ref{3Rten}) and (\ref{3Rsc}) characterize the curvature of the
observer's rest space within a general spacetime. We shall rewrite (\ref
{3Rten}) and (\ref{3Rsc}) for a magnetic universe containing a single
perfect fluid of infinite conductivity. Using (\ref{Rdsp}), (\ref{rvpr}) and
(\ref{Ray}), we obtain

\begin{eqnarray}
\mbox{}^{(3)}R_{ij} &=&\frac 23\left( \kappa \left( \mu c^2+\frac{H^2}%
2\right) -\frac{\Theta ^2}{3c^2}+\frac{\sigma ^2}{c^2}-\frac{\omega ^2}{c^2}%
+\Lambda \right) h_{ij}-\frac A{3c^2}h_{ij}+\kappa M_{ij}+\frac 1{c^4}\dot
u_i\dot u_j-  \nonumber \\
&&  \nonumber \\
&&\mbox{}\frac 1{c^2}h_i^{\hspace{1mm}k}h_j^{\hspace{1mm}q}\left(
S^{-3}\left( S^3\sigma _{kq}\right) ^{\cdot }-\dot u_{(k;q)}\right) -\frac
1{c^2}h_i^{\hspace{1mm}k}h_j^{\hspace{1mm}q}\left( S^{-3}\left( S^3\omega
_{kq}\right) ^{\cdot }-\dot u_{[k;q]}\right) .  \label{3Rten1}
\end{eqnarray}
Moreover, from (\ref{Rdtem}), (\ref{Rsc}) and (\ref{3Rsc}), we find the
following 3-curvature scalar

\begin{equation}
{\cal K}\equiv \mbox{}^{(3)}R=2\left( \kappa \left( \mu c^2+\frac{H^2}%
2\right) -\frac{\Theta ^2}{3c^2}+\frac{\sigma ^2}{c^2}-\frac{\omega ^2}{c^2}%
+\Lambda \right) .  \label{K}
\end{equation}
Combining (\ref{3Rten1}) with (\ref{K}), we can now decompose the 3-Ricci
tensor into its trace and its trace-free parts

\begin{eqnarray}
\mbox{}^{(3)}R_{ij}&=&\frac{{\cal K}}{3}h_{ij}- \frac{A}{3c^2}h_{ij}+\kappa
M_{ij}+\frac{1}{c^4}\dot{u}_i\dot{u}_j -\frac{1}{c^2}h_i^{\hspace{1mm}k}h_j^{%
\hspace{1mm}q} \left(S^{-3}\left(S^3\sigma_{kq}\right)^{\cdot}-\dot{u}%
_{(k;q)}\right)-  \nonumber \\
\nonumber \\
&\mbox{}&\frac{1}{c^2}h_i^{\hspace{1mm}k}h_j^{\hspace{1mm}q}
\left(S^{-3}\left(S^3\omega_{kq}\right)^{\cdot}-\dot{u}_{[k;q]}\right) .
\label{3Rten2}
\end{eqnarray}

The evolution of ${\cal K}$, along the observer's worldline, is governed by
the following propagation equation, derived from (\ref{K}), (\ref{Mi4}), (%
\ref{edc}) and (\ref{Ray})

\begin{equation}
\left({\cal K}-\frac{2}{c^2}\left(\sigma^2-\omega^2\right)\right)^{\cdot}= -%
\frac{2\Theta}{3}\left({\cal K}+\frac{2}{c^2}A\right)+ \frac{4\Theta}{3}%
\left(\sigma^2-\omega^2\right)- 2\kappa\sigma_{ij}M^{ij} .  \label{Kprop}
\end{equation}

The 3-curvature scalar qualifies as an additional gauge invariant variable,
provided that it is constant in the unperturbed cosmological model. In a
spatially homogeneous irrotational background, the 3-gradient ${\cal K}_i$
is always independent of the gauge choice, as is implied by the relation

\begin{equation}
{\cal K}_i\equiv \mbox{}^{(3)}\nabla_i{\cal K}= \frac{2\kappa\mu c^2}{S}
\left(D_i+\frac{1}{\kappa\mu c^2}{\cal M}_{ji}H^j\right)- \frac{4\Theta}{%
3Sc^2}{\cal Z}_i+ \frac{2}{c^2}\mbox{}^{(3)}\nabla_i
\left(\sigma^2-\omega^2\right) ,  \label{Ki}
\end{equation}
\\which will be useful in discussing the case of isocurvature perturbations.

It should be emphasized that all the relations given above refer to the
instantaneous rest space of an individual fundamental observer. For
non-rotating spacetimes, the same formulae (without the vorticity terms)
determine the geometry of the observers' common rest space.\\

\section{Approximations}

%~~~~~~~~~~~~~~~~~~~~~~~

\subsection{Linearization About a FRW universe}

%~~~~~~~~~~~~~~~~~~~~~~~~~~~~~~~~~~~~~~~~~~~~~~~

Let us assume that the background universe is described by a FRW spacetime
containing a perfect fluid and a weak large-scale magnetic field (i.e. $%
H^2\ll \mu c^2$). The gauge-invariant variables are the shear ($\sigma _{ij}$%
), the vorticity ($\omega _{ij}$), the acceleration ($\dot u_i$), the
divergence of the acceleration ($A$), its spatial gradient ($A_i$), the
electric and magnetic parts of the Weyl tensor ($E_{ij}$, $H_{ij}$), the
spatial gradients of the energy density ($X_i$, $D_i$), the spatial gradient
of the pressure ($Y_i$), the spatial gradient of the expansion ($Z_i$, $%
{\cal Z}_i$) and of the magnetic field (${\cal H}_{ij}$, ${\cal M}_{ij}$),
together with the spatial gradient (${\cal K}_i$) of the 3-curvature scalar.
All these quantities vanish in a Friedmann universe. According to (\ref
{Kprop}), the 3-curvature scalar is independent of the gauge choice,
provided that it vanishes in the background spacetime.\\ 

The formulae above are the exact, fully non-linear, propagation equations.
We shall linearize them about the FRW model. In the process we shall retain
the mass density ($\mu $), the pressure ($p$), the expansion ($\Theta $) and
the magnetic field ($H_i$), which are all spatially independent in the
background, as zero-order quantities. Variables that vanish in the FRW
background, together with their derivatives, will be treated as first order.
The anisotropic pressure ($M_{ij}$) generated by the primordial field will
be also considered as of order one, since it must disappear within a FRW
spacetime. Finally, terms of order higher than the first will be
disregarded. Under these conditions, the conservation law (\ref{edc}) for
the energy-density remains unaffected, whereas law (\ref{mdc1}), for the
momentum-density, changes. In particular, we have

\begin{equation}
\dot{\mu}c^2+\left(\mu c^2+p\right)\Theta=0 ,  \label{ledc}
\end{equation}
\\and 
\begin{equation}
\kappa\left(\mu+\frac{p}{c^2}\right)\dot{u}_i+ Y_i- 2{\cal H}_{[ij]}H^j=0 .
\label{lmdc}
\end{equation}

Raychaudhuri's equation, (\ref{Ray}), becomes

\begin{equation}
\dot \Theta -A+\frac{\Theta ^2}3+\frac{\kappa c^2}2\left( \mu c^2+3p\right)
-\Lambda c^2=0,  \label{lRay}
\end{equation}
and the  time-evolution formula for the vorticity vector, (\ref{vorpr}),
reduces to

\begin{equation}
\dot{\omega}^i+ \frac{2\Theta}{3}\omega^i= \frac{1}{2c}\eta^{ijkq}u_j\dot{u}%
_{k;q} .  \label{lvorpr}
\end{equation}

Linearization permits the omission of the projection tensor from the
left-hand side of the remaining propagation equations and the weakness of
the field implies that $D_i/2+{\cal M}_{ji}H^j/\kappa \mu c^2\simeq D_i/2$.
Hence, equations (\ref{Dprop}), (\ref{cZprop}) and (\ref{Mprop}), which
describe the evolution of $D_i$, ${\cal Z}_i$ and ${\cal M}_{ij}$
respectively, become

\begin{equation}
\dot{D}_i= \frac{p\Theta}{\mu c^2}D_i- \left(1+\frac{p}{\mu c^2}\right){\cal %
Z}_i- \frac{2\Theta}{\kappa\mu c^2}{\cal M}_{[ij]}H^j+ \frac{2S\Theta H^2}{%
3\mu c^4}\dot{u}_i ,  \label{lDprop}
\end{equation}
\\

\begin{equation}
\dot{{\cal Z}}_i= -\frac{2\Theta}{3}{\cal Z}_i- \frac{\kappa\mu c^4}{2}D_i-
3c^2{\cal M}_{[ij]}H^j+ S{\cal R}\dot{u}_i+ SA_i ,  \label{lcZprop}
\end{equation}
\\where now ${\cal R}={\cal K}/2$, with ${\cal K}=2\left(\kappa\mu
c^2-\Theta^2/3c^2+\Lambda\right)$, and

\begin{eqnarray}
\dot{{\cal M}_{ij}}&=& -\frac{2\Theta}{3}{\cal M}_{ij}- \frac{2\kappa}{3}H_i%
{\cal Z}_j+ \kappa SH^k\mbox{}^{(3)}\nabla_j\left(\sigma_{ik}+\omega_{ik}%
\right)- \frac{\kappa\Theta S}{3c^2}\left(2H_i\dot{u}_j+\dot{u}_i H_j\right)+
\nonumber \\
&\mbox{}&\frac{\kappa\Theta S}{3c^2}\dot{u^k}H_k h_{ij}- \kappa Sh_i^{%
\hspace{1mm}k}R_{kqjs}H^q u^s .  \label{lMprop}
\end{eqnarray}

Linearization leaves Maxwell's equations unaffected: hence,

\begin{equation}
H^i_{\hspace{1mm};i}-\frac{1}{c^2}H^i\dot{u_i}=0 ,  \label{lMi3}
\end{equation}
\\and

\begin{equation}
\left(\sigma^i_{\hspace{1mm}j}+\omega^i_{\hspace{1mm}j} -\frac{2\Theta}{3}%
h^i_{\hspace{1mm}j}\right)H^j= h^i_{\hspace{1mm}j}\dot{H}^j .  \label{lMi4}
\end{equation}

The key role played by the acceleration ($\dot u_i$), by its divergence ($%
A=\dot u_{\hspace{1mm};i}^i$) and by the latter's spatial gradient ($A_i=%
\mbox{}^{(3)}\nabla _iA$) will become clear below. Therefore, it is useful
to derive expressions for these three variables. The momentum-density
conservation law (\ref{lmdc}), immediately provides a useful formula for $%
\dot u_i$, that leads to the following expression for A

\begin{equation}
A=-\frac 1{\kappa \left( \mu +p/c^2\right) }\left( \mbox{}^{(3)}\nabla _iY^i-%
\frac{\kappa H^2{\cal K}}3+\frac{\kappa \mbox{}^{(3)}\nabla ^2H^2}2\right) .
\label{lA}
\end{equation}
\\ This is derived by applying the commutator (\ref{3Rid}) to the spacelike
vector $H_i$, and substituting the zero-order part of the Ricci tensor
associated with $\Sigma _{\perp }$ from (\ref{3Rten2}). Furthermore, the
spatial derivative of (\ref{lA}), combined with (\ref{Mi4}), (\ref{edc}), (%
\ref{3Rid}) and (\ref{3sc}) provides an expression for $A_i$ within a
perturbed FRW universe:

\begin{eqnarray}
A_i &=&\frac 1{\kappa \left( \mu +p/c^2\right) }\left( \frac{{\cal K}}3-%
\mbox{}^{(3)}\nabla ^2\right) Y_i+\frac 1{2\left( \mu +p/c^2\right) }\left( 
{\cal K}-\mbox{}^{(3)}\nabla ^2\right) B_i+  \nonumber \\
&&\mbox{}\frac{H^2}{3\left( \mu +p/c^2\right) }{\cal K}_i-2\Theta \left( 
\frac{c_s^2}{c^2}+\frac{2H^2}{3\left( \mu c^2+p\right) }\right) \mbox{}%
^{(3)}\nabla _j\omega _i^{\hspace{1mm}j},  \label{lAi}
\end{eqnarray}
\\ where $c_s^2\equiv \dot p/\dot \mu $ is the sound speed and $B_i=\mbox{}%
^{(3)}\nabla _iH^2$ is an additional first-order variable. 

We make one further comment about the 3-curvature scalar ${\cal K}$. Its
linearized propagation formula,

\begin{equation}
\dot {{\cal K}}=-\frac{2\Theta }3\left( {\cal K}+\frac 2{c^2}A\right) ,
\label{lKprop}
\end{equation}
\\ contains a first-order term generated by the divergence of the
acceleration. Thus, to zero-order, we find

\begin{equation}
{\cal K}=\frac{6k}{S^2},  \label{zlKprop}
\end{equation}
\\ where $k$ is the 3-curvature constant (i.e. $\dot k=0$) of the background
FRW spacetime~\cite{EB}. One can replace ${\cal K}$ by (\ref{zlKprop})
whenever the 3-Ricci scalar is coupled to a quantity of first or higher
order. Such a substitution further simplifies equations (\ref{lcZprop}) and
(\ref{lAi}). We shall adopt this approach in what follows. Finally, equation
(\ref{Ki}) reduces to

\begin{equation}
{\cal K}_i= \frac{2\kappa\mu c^2}{S}D_i- \frac{4\Theta}{3Sc^2}{\cal Z}_i ,
\label{lKi}
\end{equation}
\\which can be used to modify (\ref{lAi}), if required.

\subsection{Pressureless Fluid}

%~~~~~~~~~~~~~~~~~~~~~~~~~~~~~~
Let us assume that the perturbed FRW universe, described in the last
section, is dominated by a single non-relativistic (i.e. $p=0$) perfect
fluid. Under this restriction the equation of continuity, (\ref{ledc}),
becomes

\begin{equation}
\dot{\mu}+\mu\Theta=0 ,  \label{pledc}
\end{equation}
\\and leads to the familiar evolution formula for the mass density of a
dust-dominated cosmological model

\begin{equation}
\mu=\frac{M}{S^3} ,  \label{plmupr}
\end{equation}
\\with $\dot{M}=0$.

An important result emerges from the conservation of the momentum-density.
In particular, the contraction of the pressure-free form of (\ref{lmdc})
with the magnetic-field vector provides the relation

\begin{equation}
\dot {u_i}H^i=0.  \label{perp1}
\end{equation}
\\ Hence, within a magnetized dust-dominated nearly FRW universe, the
acceleration must always be normal to the magnetic field, a result
consistent with the Lorentz force law of special relativity. Furthermore,
equation (\ref{perp1}) may be rearranged to give

\begin{equation}
\dot H_iu^i=0,  \label{perp2}
\end{equation}
\\ and ensures that the field's derivative actually lies in the observer's
rest space. Also,  $p=0$ in (\ref{lmdc}) leads to the following expression
for the acceleration

\begin{equation}
\dot {u_i}=\frac 2{\kappa \mu }{\cal H}_{[ij]}H^j.  \label{pacc}
\end{equation}
\\ This demonstrates how the geodesic motion of the matter is disturbed by
the spatial variations of the magnetic field. We can express (\ref{pacc}) in
a more familiar form: ${\cal H}_{[ij]}$ is an antisymmetric 3-tensor, the
generalized curl of $H_i$ projected onto the observer's rest space. Such a
tensor has only three independent components and corresponds to a 3-vector,
namely the covariant spatial curl of $H_i$. In particular (see Appendix
C.1), we find

\begin{equation}
{\cal H}_{[ij]}=-\frac{\kappa}{2}\epsilon_{ijk}curlH^k ,  \label{H[ij]}
\end{equation}
\\where $\epsilon_{ijk}=\eta_{ijkq}u^q /c$ is the covariant spatial
permutation tensor. When substituted into (\ref{pacc}), (\ref{H[ij]})
provides the following expression for the acceleration, familiar from
classical magnetohydrodynamics

\begin{equation}
\dot{u}_i=\frac{1}{\mu}\epsilon_{ijk}((curlH^j)H^k) .  \label{pacc1}
\end{equation}
\\Notice that results (\ref{perp1}), (\ref{perp2}) and (\ref{pacc}), or
equivalently (\ref{pacc1}), also hold within a pressure-free almost
Bianchi-I universe.

An additional familiar result is obtained when we adapt the vorticity
propagation formula, (\ref{vorpr}), to our cosmological model. From $%
\omega^i=-curlu^i /2$ and equation (\ref{pacc1}), we find that

\begin{equation}
(curlu^i)^{{\bf .}}+ \frac{2\Theta}{3}curlu^i= \frac{1}{\mu}\epsilon^{ijk}%
\mbox{}^{(3)}\nabla_j \left(\epsilon_{kqs}(curlH^q)H^s\right) ,
\label{curlupr}
\end{equation}
\\in agreement with Wasserman's analysis~\cite{W}.

The adaptation of Maxwell's equations to this cosmological model gives

\begin{equation}
H^i_{\hspace{1mm};i}=0 ,  \label{plMi3}
\end{equation}
\\and

\begin{equation}
\left( \sigma _{\hspace{1mm}j}^i+\omega _{\hspace{1mm}j}^i-\frac{2\Theta }%
3h_{\hspace{1mm}j}^i\right) H^j=\dot H^i,  \label{plMi4}
\end{equation}
\\ with the latter providing the standard expression $H^2\propto S^{-4}$ for
the evolution of the energy density of the magnetic field with the expansion
scale factor $S(t)$.

We shall now proceed with some additional simplifications. Suppose that
spacetime is flat (i.e. $R_{ijkq}=0$) and that there is no cosmological
constant (i.e. $\Lambda =0$), then Raychaudhuri's equation (\ref{lRay})
reduces to

\begin{equation}
\dot \Theta =A+\frac{{\cal K}c^2}2-\frac{3\kappa \mu c^4}2,  \label{plRay}
\end{equation}
\\ since ${\cal K}=2\left( \kappa \mu c^2-\Theta ^2/3c^2\right) $.Using (\ref
{perp1}) and (\ref{pacc}), the propagation equations for the spatial
gradients (\ref{lDprop}), (\ref{lcZprop}) and (\ref{lMprop}) become

\begin{eqnarray}
\dot{D}_i&=& -{\cal Z}_i- \frac{2\Theta}{\kappa\mu c^2}{\cal M}_{[ij]}H^j ,
\label{plDprop} \\
\nonumber \\
\dot{{\cal Z}}_i&=& -\frac{2\Theta}{3}{\cal Z}_i- \frac{\kappa\mu c^4}{2}%
D_i- 3c^2{\cal M}_{[ij]}H^j+ SA_i+ \frac{6k}{S^2\kappa\mu}{\cal M}_{[ij]}H^j
,  \label{plcZprop} \\
\nonumber \\
\dot{{\cal M}_{ij}}&=& -\frac{2\Theta}{3}{\cal M}_{ij}- \frac{2\kappa}{3}H_i%
{\cal Z}_j+ \kappa SH^k\mbox{}^{(3)}\nabla_j\left(\sigma_{ik}+\omega_{ik}%
\right)+ \frac{2\Theta H^2}{9\mu c^2}{\cal M}_{[ij]} ,  \label{plMprop}
\end{eqnarray}
\\respectively, where the pressure-free expression for the quantity $A_i$,
that appears in (\ref{plcZprop}), is

\begin{equation}
A_i= \frac{1}{2\mu}\left(\frac{6k}{S^2}-\mbox{}^{(3)}\nabla^2\right)B_i+ 
\frac{H^2}{3\mu}{\cal K}_i- \frac{4\Theta H^2}{3\mu c^2}\mbox{}%
^{(3)}\nabla_j\omega_i^{\hspace{1mm}j} ,  \label{plAi}
\end{equation}
\\$B_i=\mbox{}^{(3)}\nabla_i H^2$, and ${\cal K}_i$ is given by (\ref{lKi}).

The antisymmetry of ${\cal M}_{[ij]}$ means that the indices $i$ and $j$ in (%
\ref{plDprop}) must take different values ($i\neq j$). Thus, irregularities
in the energy density are not influenced by the component of the field that
acts parallel to them, a result which is also apparent in the analysis given
by Ruzmaikina and Ruzmaikin~\cite{RR}. Combining equations (\ref{plDprop})-(%
\ref{plAi}) and ignoring the non-linear terms, we obtain the following
second-order differential equation for the time-evolution of the comoving
fractional orthogonal spatial gradient of the energy density,

\begin{eqnarray}
\ddot D_i &=&-\frac{2\Theta }3\dot D_i+\frac{\kappa \mu c^4}2D_i-\frac
S{2\mu }\left( \frac{6k}{S^2}-\mbox{}^{(3)}\nabla ^2\right) B_i+\frac{6k}{%
S^2\kappa \mu }{\cal M}_{[ij]}H^j+  \nonumber \\
&&\mbox{}\frac{4\Theta SH^2}{3\mu c^2}\mbox{}^{(3)}\nabla _j\omega _i^{%
\hspace{1mm}j}-\frac{2\Theta S}{\mu c^2}\mbox{}^{(3)}\nabla _{[j}\dot
H_{i]}H^j.  \label{plDprop1}
\end{eqnarray}
\\ In a perturbed Einstein de Sitter universe (i.e. $k=0$ and $\Lambda =0$)
this reduces to

\begin{equation}
\ddot{D}_i= -\frac{2\Theta}{3}\dot{D}_i+ \frac{\kappa\mu c^2}{2}D_i+ \frac{S%
}{2\mu}\mbox{}^{(3)}\nabla^2 B_i+ \frac{4\Theta SH^2}{3\mu c^2}\mbox{}%
^{(3)}\nabla_j \omega_i^{\hspace{1mm}j}- \frac{2\Theta S}{\mu c^2}\mbox{}%
^{(3)}\nabla_{[j}\dot{H}_{i]}H^j  \label{plDprop2}
\end{equation}

The first three terms on the right hand side of (\ref{plDprop2}) also appear
in the Newtonian treatments presented by Ruzmaikina and Ruzmaikin \cite{RR},
and by Wasserman \cite{W}; the first two are immediately identifiable, while
the third is obtained after a transformation (see Appendix C.2). The latter
becomes obvious, when the gauge-invariant approach is applied within the
non-relativistic context, as it has been shown by Ellis in \cite{E3}. The
resulting second-order differential equation contains only the three
aforementioned standard terms. Consequently, our relativistic approach
provides the classical equation plus two extra relativistic terms. The first
term links the vorticity to the time evolution of linear density gradients
through the 3-divergence of the vorticity tensor. It appears because in a
general spacetime 3-D surfaces need not be orthogonal to the fluid-flow.
However, the term's coefficient suggests a weak effect since $H^2\ll \mu c^2$%
, or no effect at all if the universe is static. Such a `correction term'
does not appear in the Newtonian treatment, because there the hyperplanes
orthogonal to the fluid flow are always tangent to the 3-surfaces of
absolute time \cite{EBH}. Also, using (\ref{vt}) and the covariant
definition of a vector's spatial curl we find

\begin{equation}
\frac{4\Theta SH^2}{3\mu c^2}\mbox{}^{(3)}\nabla _j\omega _i^{\hspace{1mm}j}=%
\frac{4\Theta SH^2}{3\mu c^2}curl\omega _i,  \label{ct1}
\end{equation}
\\ where $\omega ^i=-curlu^i/2$.\\ Let us focus on the last term in (\ref
{plDprop2}). It emerges from the coupling between $D_i$ and ${\cal M}_{ij}$
in (\ref{Dprop}). We see that its presence also depends on the universal
expansion, but it also creates only a weak effect, since it contains the
ratio $H^2/\mu c^2\ll 1$. Similarly to the acceleration case above, the
quantity $\mbox{}^{(3)}\nabla_{[j}\dot H_{i]}$ represents the generalized curl
of the field's derivative projected onto the instantaneous rest space of a
fundamental observer. Thus, it can be mapped onto the covariant spatial curl
of $\dot H_i$. As a result, the second relativistic `correction term' can be
reshaped into the following easily identifiable form

\begin{equation}
\frac{2\Theta S}{\mu c^2}\mbox{}^{(3)}\nabla_{[j}\dot{H}_{i]}H^j= \frac{%
\Theta S}{\mu c^2}\epsilon_{ijk}(curl\dot{H}^j)H^k .  \label{ct2}
\end{equation}
\\Thus, according to (\ref{ct1}) and (\ref{ct2}), equation (\ref{plDprop2})
becomes

\begin{equation}
\ddot{D}_i= -\frac{2\Theta}{3}\dot{D}_i+ \frac{\kappa\mu c^2}{2}D_i+ \frac{S%
}{2\mu}\mbox{}^{(3)}\nabla^2 B_i+ \frac{4\Theta SH^2}{3\mu c^2}curl\omega_i- 
\frac{\Theta S}{\mu c^2}\epsilon_{ijk}(curl\dot{H}^j)H^k .  \label{plDprop3}
\end{equation}
\\

\subsection{Long-Wavelength Solutions}

%~~~~~~~~~~~~~~~~~~~~~~~~~~~~~~~~~~~~
The gradient $D_i$ describes spatial variations of the energy density
orthogonal to the fluid flow. One can extract the information contained in $%
D_i$ by considering the following local decomposition \cite{EBH},

\begin{equation}
S\mbox{}^{(3)}\nabla _iD_j\equiv \Delta _{ij}=\Sigma _{ij}+W_{ij}+\frac
13\Delta h_{ij},  \label{Delij}
\end{equation}
\\ where $\Sigma _{ij}\equiv \Delta _{(ij)}-\Delta h_{ij}/3$, $W_{ij}\equiv
\Delta _{[ij]}$ and $\Delta \equiv S\mbox{}^{(3)}\nabla ^iD_i$ are all
gauge-invariant quantities. The symmetric, traceless tensor, $\Sigma _{ij}$,
describes the evolution of anisotropic structures (e.g. pancake or
cigar-like structures). On the other hand, the skew tensor $W_{ij}$
characterizes the rotational behaviour of $D_i$. Hence, the scalar $\Delta $
contains the information regarding the spatial aggregation of matter. When
dealing with the problem of structure formation, the latter quantity is the
crucial one \cite{BDE}, \cite{DBE}. The propagation formula for $\Delta $ is
found by taking the 3-divergence of (\ref{plDprop1}) and using the
properties of the 3-D gradients as well as the fact that, in the linear
approximation, the total divergence of the vorticity tensor is zero (see
section 7 and also~\cite{EBH}). We obtain

\begin{equation}
\ddot \Delta +\frac{2\Theta }3\dot \Delta -\frac{\kappa \mu c^4}2\Delta=
-\frac{S^2}{2\mu }\left(\frac{4k}{S^2}-\mbox{}^{(3)}\nabla ^2\right){\cal B}
+\frac{S^2}{2\mu}\left(\frac{3k}{S^2}+\frac{2\Theta^2}{3c^2}\right)
\left(\frac{4k}{S^2}-\mbox{}^{(3)}\nabla^2\right)H^2
\label{Delprop}
\end{equation}\\
where ${\cal B}=\mbox{}^{(3)}\nabla ^iB_i=\mbox{}^{(3)}\nabla ^2H^2$. The
final form of (\ref{Delprop}) is obtained only after treating the field's
energy density ($H^2$) as a perturbation itself.\footnote{%
Although it has not been explicitly used yet, the characterization of $H^2$
as a gauge-invariant perturbative term is in line with our treatment of $%
M_{ij}$. Nevertheless, up to this point, we have only used the restriction
that the ratio $H^2/\mu c^2$ is very small.}

Assuming that spatial and temporal dependences are separable, we express
every gauge-invariant first-order quantity in (\ref{Delprop}) as a sum of
time-independent scalar harmonics $Q^{(n)}$, defined as eigenfunctions of
the Laplace-Beltrami operator, \cite{H},

\begin{equation}
\mbox{}^{(3)}\nabla ^2Q^{(n)}=-\frac{n^2}{S^2}Q^{(n)},  \label{sh}
\end{equation}
where the eigenvalue $n$ directly corresponds to physical wavelengths only
if $k=0$ \cite{DBE}. Thus, in an almost Einstein de Sitter universe, and for
low frequency (i.e. $n\rightarrow 0$) fluctuations, we may ignore all the
terms in the right-hand side of (\ref{Delprop}). Consequently, the
large-scale evolution of matter perturbations is determined by

\begin{equation}
\ddot{\Delta}^{(n)}+ \frac{2\Theta}{3}\dot{\Delta}^{(n)}- \frac{\kappa\mu c^4%
}{2}\Delta^{(n)}=0 ,  \label{Delprop1}
\end{equation}
\\where $\Delta^{(n)}$ is the nth harmonic component of $\Delta$, with $%
\mbox{}^{(3)}\nabla_i \Delta^{(n)}\approx 0$. Its solution,

\begin{equation}
\Delta ^{(n)}=\Delta _{+}^{(n)}\tau ^{2/3}+\Delta _{-}^{(n)}\tau ^{-1},
\label{Delsol}
\end{equation}
contains a growing and decaying modes, exactly as in the case of a
dust-dominated non-magnetized universe and in agreement with~\cite{RR}. The
variable $\tau $ measures the observer's proper time, while the quantities $%
\Delta _{+}^{(n)}$ and $\Delta _{-}^{(n)}$ remain constant along its
worldline.\\ 

\section{Two Special Cases}

%~~~~~~~~~~~~~~~~~~~~~~~~~~

\subsection{The False Vacuum Assumption}

%~~~~~~~~~~~~~~~~~~~~~~~~~~~~~~~~~~~~~~~
The introduction of a medium that obeys the ``inflationary'' false vacuum
condition (i.e. $p+\mu c^2=0$), implies that the spatial gradients $X_i$ and 
$Y_i$, defined by (\ref{X})-(\ref{Y}), are related by the simple formula

\begin{equation}
Y_i=-X_i .  \label{YiXi}
\end{equation}

The exact propagation equation for the orthogonal spatial gradient of the
energy density, $X_i$, is obtained by taking the time derivative of (\ref{X}%
). From the conservation laws (\ref{edc}) and (\ref{mdc1}), we have

\begin{eqnarray}
S^{-4}h_i^{\hspace{1mm}j}\left( S^4X_j\right) ^{.} &=&-X^j\left( \sigma _{%
\hspace{1mm}i}^j+\omega _{\hspace{1mm}i}^j\right) -\kappa \left( \mu
c^2+p\right) Z_i-2\Theta {\cal H}_{[ij]}H^j+  \nonumber \\
&&\mbox{}\frac{2\kappa \Theta H^2}{3c^2}\dot u_i+\frac{\kappa \Theta }{c^2}%
\dot u^jM_{ji},  \label{Xprop}
\end{eqnarray}
\\ where $Z_i$ has been defined in (\ref{Z}). In an almost-FRW magnetized
universe, the false vacuum assumption, together with the momentum-density
conservation law, reduces (\ref{Xprop}) to the following evolution formula

\begin{equation}
X_i=\frac XS,  \label{Xev}
\end{equation}
\\ with $\dot X=0$. Hence, spatial gradients in the energy density die away
as $S^{-1}$, independent of their wavelength and in line with the cosmic
``no-hair'' theorems for expanding inflationary universes.

According to (\ref{edc}), however, the energy density of the material
component remains constant along the observers worldline. Consequently,
definition (\ref{D}), along with (\ref{Xev}), suggests that the fractional
density gradients do not change on comoving scales

\begin{equation}
D_i=const.  \label{Dev}
\end{equation}
\\ In other words, fundamental observers experience no variation in $D_i$
within their event horizons: density contrasts remain ``frozen-in'' as long
as the equation of state $p=-\mu c^2$ holds.\\ 

\subsection{Isocurvature Perturbations}

%~~~~~~~~~~~~~~~~~~~~~~~~~~~~~~~~~~~~~~
We define isocurvature perturbations as fluctuations evolving in a
non-rotating (i.e. $\omega _{ij}=0$) spacetime of constant spatial curvature
(i.e. ${\cal K}_i=0$). Within an almost-FRW universe, equation (\ref{lKi})
provides a condition for the occurrence of isocurvature inhomogeneities:

\begin{equation}
X_i=\frac{2\Theta}{3c^2}Z_i .  \label{ic}
\end{equation}

The consistency condition for this kind of disturbance is obtained by
linearizing the time derivative of (\ref{ic}) about the background FRW
spacetime, giving

\begin{equation}
\frac{6k}{S^2}\left( Z_i+\frac \Theta {c^2}\dot {u_i}\right) +\frac{2\Theta 
}{c^2}A_i=0,  \label{cic}
\end{equation}
\\ where $A_i$ is given by (\ref{lAi}).

Provided that the smooth model has $k=0$, we can ignore the first term in (%
\ref{cic}), which means that isocurvature disturbances cannot survive in a
non-static universe so long as $A_i\neq 0$. On the other hand, in the
absence of pressure, formula (\ref{lAi}) implies that the remaining term is
unimportant on large scales. Consequently, the matter era seems capable of
preserving long-wavelength isocurvature fluctuations.\\ 

\section{Discussion}

%~~~~~~~~~~~~~~~~~~~~~
We have pursued the study of cosmological density perturbations in a
universe that contains a large-scale primordial magnetic field by means of
the Ellis-Bruni covariant and gauge-invariant method. We have defined a new
variable that is independent of the gauge choice and describes the spatial
variations of the magnetic field. Assuming a universe that contains a single
perfect fluid of very high conductivity, we have derived  the exact, and
fully non-linear, equations that determine the model's time evolution. When
the linearized equations are applied to the simple case of pressure-free
matter, familiar results are recovered. The energy density of the matter is
found to evolve as the inverse cube of the scale factor and the energy
density of the magnetic field as $S^{-4}$. The acceleration is provided by
the well known expression of classical magnetohydrodynamics, and is always
normal to the magnetic field in accord with the Lorentz force law. 

We also identify the general-relativistic corrections to the Newtonian
treatment. These are introduced by the two extra terms (\ref{ct1}) and (\ref
{ct2}) on the right-hand side of the propagation equation, (\ref{plDprop2}),
of the density gradient. The first term is due to universal rotation and
does not affect the gravitational clumping of matter. A similar vorticity
term also appears in the treatment of a perfect fluid with non-vanishing
pressure, \cite{EBH}. Here the fluid pressure is zero, but there exists a
residual isotropic pressure, induced into the model by the magnetic field
(see (\ref{cor2})). Technically, the second relativistic correction, which
modifies the evolution of both $D_i$ and $\Delta $, appears because
everything is projected onto the observer's instantaneous rest space (where
all measurements take place). Physically, its source  is the anisotropic
pressure, also introduced by the field (see (\ref{cor3})). In Dunsby's
imperfect fluid analysis, \cite{D1}, such a term is incorporated into the
time derivative $(\mbox{}^{(3)}\nabla _j\pi _i^{\hspace{1mm}j})_{;k}u^k$.

By focussing upon the spatial aggregation of matter, we have obtained
long-wavelength solutions for (\ref{plDprop1}), which reveal the relative
unimportance of the magnetic field on the growth of large-scale density
fluctuations. We also provide solutions during a period of de Sitter
inflation. The results show that density gradients decay, but slower than in
the non-magnetized case investigated by Ellis and Bruni, \cite{EB}.
Furthermore, within the immediate neighbourhood of a comoving observer
perturbations freeze out. Our analysis permits a transparent approach to
isocurvature inhomogeneities. We have obtained a criterion for the
occurrence of this kind of disturbances as well as a consistency condition.
The latter shows that a dust-dominated almost-FRW universe, can sustain
long-wavelength isocurvature perturbations.

A major advantage of the Ellis and Bruni technique is that the non-linear
propagation equations can be often linearized about a variety of background
models. For example, we can linearize our exact equations about a smooth
Bianchi-I universe. This cosmological model is anisotropic. The shear no
longer behaves as a first-order variable and the model's evolution becomes
more complicated, \cite{Ba}. The formalism we have established here, will
enable a full gauge-invariant analysis to be carried out to determine the
effects of cosmological magnetic and electric fields on the early universe,
and on the temperature anisotropy of the microwave background.\\ 

\section*{Acknowledgements}

%~~~~~~~~~~~~~~~~~~~~~~~~~
C.G.T. is supported by the Greek State Scholarship Foundation and J.D.B. by
the PPARC. The authors wish to thank V.G. Gurzadyan, K. Subramanian and
particularly R. Maartens for helpful discussions.\\

\section*{APPENDICES}

%~~~~~~~~~~~~~~~~~~~~
\appendix

\section{The Spatial Gradient of the Magnetic Field}

%~~~~~~~~~~~~~~~~~~~~~~~~~~~~~~~~~~~~~~~~~~~~~~~~~~~

\subsection{Proof of the Gauge-Invariance for ${\cal M}_{ij}$}

%~~~~~~~~~~~~~~~~~~~~~~~~~~~~~~~~~~~~~~~~~~~~~~~~~~~~~~~~~~~~~

The metric associated with an exact FRW or with an exact Bianchi-I spacetime
can always be written in the diagonal form $%
g_{ij}=diag(g_{00},g_{11},g_{22},g_{33})$ with the quantities $g_{ii}$ (no
sum) functions of time only (i.e. $g_{ij,\mu }=0$).\footnote{%
Latin indices run from 0 to 3, whereas Greek ones from 1 to 3.} As a result
the associated spatial Christoffel symbols vanish

\begin{equation}
\Gamma^\mu_{\hspace{1mm}\nu\kappa}=0 .  \label{Gamma}
\end{equation}
\\Following definition (\ref{Hten}), we split the spatial gradient of the
magnetic field as

\begin{eqnarray}
{\cal H}_{ij} &=&\kappa h_i^{\hspace{1mm}\mu }h_j^{\hspace{1mm}\nu }\left(
H_{\mu ,\nu }-\Gamma _{\hspace{1mm}\mu \nu }^\kappa H_\kappa \right) -\kappa
h_i^{\hspace{1mm}\mu }h_j^{\hspace{1mm}\nu }\Gamma _{\hspace{1mm}\mu \nu
}^0H_0+  \nonumber \\
&&\mbox{}\kappa h_i^{\hspace{1mm}\mu }h_j^{\hspace{1mm}0}H_{\mu ;0}+\kappa
h_i^{\hspace{1mm}0}h_j^{\hspace{1mm}q}H_{0;q}.  \label{dHij}
\end{eqnarray}
\\ In a spatially homogeneous universe $H_{i,\mu }=0$. Also, in a comoving
frame, equation (\ref{P2}) means that $H_0=0$. Also, when there is no
rotation, $u_i=-c\delta _i^{\hspace{1mm}0}$, so $h_i^{\hspace{1mm}0}=0$.
Thus, via (\ref{Gamma}), relative to a comoving frame, equation (\ref{dHij})
gives

\begin{equation}
{\cal H}_{ij}=0.
\end{equation}
\\ Therefore, definition (\ref{Hten}) provides a spatial gradient for the
magnetic field that satisfies the requirements for gauge invariance, when
the background universe is a FRW or a Bianchi-I spacetime.\\ 

\subsection{Description of Magnetic Spatial Variations by ${\cal M}_{ij}$}

%~~~~~~~~~~~~~~~~~~~~~~~~~~~~~~~~~~~~~~~~~~~~~~~~~~~~~~~~~~~~~~~~~~~~~~~~~~~~
Let us consider two neighbouring fundamental observers $P$ and $P^{\prime }$%
; \{$x^i$\} is a coordinate system comoving with the expanding fluid.
Relative to this frame, the timelike worldlines associated with the two
observers, which are assumed to be close enough to measure the same proper
time ($\tau $), are labelled by ($x^\mu =const,c\tau $) and by ($x^i+\delta
x^i,c\tau $) respectively. With respect to a general frame $y^i=y^i(x^j)$,
the vector $(\delta x^\mu ,0)$, which connects the two worldlines at all
times, becomes

\begin{equation}
\delta y^i=\frac{\partial y^i}{\partial x^\nu}\delta x^\nu .  \label{dy}
\end{equation}
\\As far as the observer $P$ is concerned, the relative position vector
between the two events is the projection of $\delta y^i$ onto its own 3-D
instantaneous rest space $\Sigma_{\perp}$

\begin{equation}
\delta_{\perp}y^i=h^i_{\hspace{1mm}j}\delta y^j .  \label{pdy}
\end{equation}
\\In order to examine the field's spatial variation, the observer at $P$
must parallel transport its own vector $\left(H_i\right)_P$ to $P^{\prime}$
and compare it to the one defined there by the field itself. The resulting
difference should then be projected onto $\Sigma_{\perp}$. The field vector
at $P^{\prime}$ as seen from $P$ is

\begin{equation}
\left(H_i\right)_{P^{\prime}}= H_i\left(y^j+\delta_{\perp}y^j\right)=
\left(H_i\right)_P+\left(H_{i,j}\right)_P\delta_{\perp}y^j .  \label{HP'}
\end{equation}
\\The parallel transport of $\left(H_i\right)_P$ to $P^{\prime}$, along the
infinitesimal displacement $\delta_{\perp}y^i$, generates the vector

\begin{equation}
\left(H_i^*\right)_{P^{\prime}}= \left(H_i\right)_P+ \left(\Gamma^k_{%
\hspace{1mm}ij}\right)_P\left(H_k\right)_P\delta_{\perp}y^j .  \label{ptHP}
\end{equation}
\\The spacelike part of the difference between (\ref{HP'}) and (\ref{ptHP})
gives the variation of the magnetic field at $P^{\prime}$ as measured by the
observer at $P$. Since $P$ is arbitrary, we have

\begin{eqnarray}
\delta_{\perp}H_i&=& h_i^{\hspace{1mm}j}\left(H_{j,k}-\Gamma^q_{jk}H_q%
\right)\delta_{\perp}y^k  \nonumber \\
\mbox{}&=&\frac{1}{\kappa}{\cal H}_{ij}\delta_{\perp}y^j .  \label{dH}
\end{eqnarray}
\\Within an almost FRW spacetime, the relative position vector evolves as $%
\delta_{\perp}y^i=S\left(\delta_{\perp}y^i\right)_0$, where $%
\left(\delta_{\perp}y^i\right)_0=const$, \cite{EB}. Hence, we arrive at the
formula

\begin{equation}
\delta _{\perp }H_i=\frac 1\kappa S{\cal H}_{ij}\left( \delta _{\perp
}y^j\right) _0,  \label{dH1}
\end{equation}
\\ and conclude that the second-order tensor ${\cal M}_{ij}=S{\cal H}_{ij}$
determines the spatial variations of the magnetic field between two
neighbouring fundamental observers.\\ 

\section{The Case of a Pure Magnetic Field}

%~~~~~~~~~~~~~~~~~~~~~~~~~~~~~~~~~~~~~~~~~~
The adoption of an infinitely conducting medium has led to the omission of
the electric field from Maxwell's equations, while preserving the spatial
currents. An alternative approach is to assume a pure magnetic field with no
electric field and zero currents. This assumption reduces (\ref{M2}) to

\begin{equation}
\eta^{ijkq}u_j\left(\dot{u}_k H_q-c^2 H_{k;q}\right)=0 ,  \label{Mp2}
\end{equation}
\\and subsequently to

\begin{equation}
\dot u_{[i}H_{j]}=\frac{c^2}\kappa {\cal H}_{[ij]}.  \label{Mp21}
\end{equation}
\\ Superficially, (\ref{Mp21}) seems to provide a relation between the
acceleration and the field gradients. Yet, when inserted into (\ref{mdc1}),
it causes the magnetic terms to cancel out, thus effectively disconnecting
inhomogeneities in the field from those in the energy density of the medium.
Some coupling between $D_i$ and ${\cal M}_{ij}$ can still be achieved via
the third term in the right-hand side of (\ref{cZprop}). However, this
coupling is only significant when $H^2\sim \mu c^2$, but is negligible in
the case of a weak magnetic field discussed here.

\section{Aspects of the Pressureless Case}

%~~~~~~~~~~~~~~~~~~~~~~~~~~~~~~~~~~~~~~~~~

\subsection{An Expression for ${\bf {\cal H}_{[ij]}}$}

%~~~~~~~~~~~~~~~~~~~~~~~~~~~~~~~~~~~~~~~~~~~~~~~~~~~~~
From the definition of the covariant spatial curl of a 3-vector (see \cite{M}%
) and the total antisymmetry of $\epsilon _{ijk}$ we have

\begin{eqnarray}
curlH^i &=&\epsilon ^{ijk}\mbox{}^{(3)}\nabla _jH_k  \nonumber \\
\mbox{} &=&-\frac 1\kappa \epsilon ^{ijk}{\cal H}_{[jk]}.  \label{curlH}
\end{eqnarray}
\\ Thus, contracting with the $\epsilon $-tensor, and using the identity $%
\epsilon ^{ijk}\epsilon _{psk}=2h_{\hspace{1mm}p}^{[i}h_{\hspace{1mm}s}^{j]}$%
, we obtain

\begin{equation}
{\cal H}_{[ij]}=-\frac{\kappa}{2}\epsilon_{ijk}curlH^k.
\end{equation}
\\

\subsection{ Comparison with the Newtonian Approach}

%~~~~~~~~~~~~~~~~~~~~~~~~~~~~~~~~~~~~~~~~~~~~~~~~~~~
In Newtonian theory, and in comoving coordinates, the evolution of the
density contrast $\delta =\delta \mu /\mu $ is determined by, \cite{RR},

\begin{equation}
\frac{\partial ^2\delta }{\partial t^2}=-\frac{2\Theta }3\frac{\partial
\delta }{\partial t}+4\pi G\mu \delta +\frac{H^\mu }{\mu S^2}\nabla ^2\delta
H_\mu .  \label{Nt}
\end{equation}
\\ By contrast, the covariant gauge-invariant approach provides the
following propagation formula for the comoving fractional orthogonal spatial
energy-density gradient,

\begin{equation}
\ddot D_i=-\frac{2\Theta }3\dot D_i+\frac{\kappa \mu c^2}2D_i+\frac S{2\mu }%
\mbox{}^{(3)}\nabla ^2B_i+\frac{4\Theta SH^2}{3\mu c^2}\mbox{}^{(3)}\nabla
_j\omega _i^{\hspace{1mm}j}-\frac{2\Theta S}{\mu c^2}\mbox{}^{(3)}\nabla
_{[j}\dot H_{i]}H^j.  \label{Rl}
\end{equation}
\\ Relative to a comoving frame $\dot D=\partial D/\partial t$, \cite{E1},
so the correspondence between the first three terms in (\ref{Nt}) and (\ref
{Rl}) is straightforward, due to the established correspondence of  $D_i$
and $\delta $ (see Appendix A in \cite{EB}). Also, recalling that $B_i=%
\mbox{}^{(3)}\nabla _iH^2$, the fourth term on the right-hand side of (\ref
{Rl}) is

\begin{equation}
\frac S{2\mu }\mbox{}^{(3)}\nabla ^2B_i=\frac{H^j}{\kappa \mu S^2}\mbox{}%
^{(3)}\nabla ^2{\cal M}_{ji},  \label{plAi1}
\end{equation}
\\ in comoving coordinates. Since the tensorial quantity ${\cal M}%
_{ij}/\kappa $ describes the infinitesimal spatial variations of the
magnetic field (see appendix A.2), the direct correspondence between the
fourth terms in (\ref{Nt}) and (\ref{Rl}) is established. \\

\end{document}